\documentstyle[12pt]{article}
\begin{document}
\begin{titlepage}
\begin{centering}

\vspace{2cm}

{\Large\bf Neutrino Decay of Positronium}\\
\vspace{0.5cm}
{\Large\bf in the Standard Model and Beyond}\\

\vspace{4cm}

Jan Govaerts\footnote{E-mail address: {\tt govaerts@fynu.ucl.ac.be}}
and Marc Van Caillie\\

\vspace{0.5cm}

{\em Institut de Physique Nucl\'eaire}\\
{\em Universit\'e catholique de Louvain}\\
{\em B-1348 Louvain-la-Neuve, Belgium}\\

\vspace{2cm}

\begin{abstract}

\noindent Whether in the Standard Model
or beyond it, neutrinos contribute to the invisible decay
mode of orthopositronium but practically not at all to that of
parapositronium. Although this remark does not resolve
the orthopositronium decay puzzle,
it allows for upper bounds to be set on neutrino
magnetic moments.

\end{abstract}

\end{centering} 

\vspace{8cm}

\noindent UCL-IPN-96-P01\\
February 1996

\end{titlepage}

\setcounter{footnote}{0}

\noindent 1. Positronium decay\cite{Rich,oPS} has provided for
a number of years a somewhat uncomfortable crack in this otherwise
most beautiful edifice which is QED.
Indeed, the two most precise
experimental rates for the $C$-odd triplet orthopositronium (o-PS) 
decay\cite{Westbrook,Nico},
\begin{equation}
\lambda^{\rm exp}_{\rm T}\ =\ 7.0514\,\pm\,0.0014\ \mu{\rm s}^{-1}\ \ \ ,\ \ \ 
\lambda^{\rm exp}_{\rm T}\ =\ 7.0482\,\pm\,0.0016\ \mu{\rm s}^{-1}\ \ \ ,
\end{equation}
are in flagrant conflict with the QED prediction\cite{oPSth,Khri,oPS},
\begin{equation}
\lambda^{\rm QED}_{\rm T}({\rm PS}\rightarrow 3\gamma)\ =\ 
7.03831\,\pm\,0.00005\ \mu{\rm s}^{-1}\ \ \ .
\label{eq:oPSth}
\end{equation}
Note that the experimental results correspond, respectively, to a
$9.4\,\sigma$ and a $6.2\,\sigma$ deviation from the
theoretical expectation, and in relative terms, to the differences,
\begin{equation}
\frac{\lambda^{\rm exp}_{\rm T}-\lambda^{\rm QED}_{\rm T}}
{\lambda^{\rm QED}_{\rm T}}\ =\ 
\left(\,1.86\ {\rm or}\ 1.41\,\right)\cdot 10^{-3} \ \ \ .
\label{eq:reldisc}
\end{equation}
In contradistinction, the $C$-even singlet parapositronium (p-PS)
decay rate, of which the most precise measurement
gives\cite{AlRamadhan},
\begin{equation}
\lambda^{\rm exp}_{\rm S}\ =\ 7990.0\,\pm\,1.7\ \mu{\rm s}^{-1}\ \ \ ,
\end{equation}
compares better with the theoretical expectation\cite{pPSth,Khri},
\begin{equation}
\lambda^{\rm QED}_{\rm S}({\rm PS}\rightarrow 2\gamma)\ =\ 
\frac{1}{2}\,\frac{\alpha^5 m_ec^2}{\hbar}\,
\Big[\,1-\frac{\alpha}{\pi}\left(5-\frac{\pi^2}{4}\right)-
2\alpha^2\ln\alpha\,+\,\cdots\,\Big]\ =\ 7989.5\ \mu{\rm s}^{-1}\ \ \ .
\end{equation}
Recently however, a new measurement of the o-PS decay rate
has been published\cite{Asai}, 
\begin{equation}
\lambda^{\rm exp}_{\rm T}\ =\ 7.0398\,\pm\,0.0025({\rm stat.})\,
\pm\,0.0015({\rm syst.})\ \mu{\rm s}^{-1}\ \ \ ,
\label{eq:Asai}
\end{equation}
in good agreement with the value in
(\ref{eq:oPSth}), since it corresponds to a $0.5\,\sigma$
deviation from the theoretical prediction and a relative
difference with it of $2.12\cdot 10^{-4}$. 
Clearly, an independent experimental
confirmation of this beautiful result is desirable, before 
definitely concluding
that the orthopositronium problem was indeed an experimental one.

Over the years, many possible explanations for the discrepancy in the
o-PS decay rate have been suggested and analysed,
both theoretically and experimentally\cite{oPS}. Among these, one may mention
mirror photons\cite{Glash}, decay modes involving weakly coupled (pseudo)scalar
bosons or even so-called invisible decay modes. However, none of these
possibilities can accomodate the complete o-PS discrepancy above.
For example, (pseudo)scalar decay modes typically lead to the following
upper limit on the o-PS branching ratio\cite{oPS,Maeno},
\begin{equation}
Br({\rm o-PS}\rightarrow \gamma+X)\ 
\stackrel{<}{_\sim} 10^{-4} - 10^{-6}\ \ \ ,
\end{equation}
depending on the mass of the particle $X$,
while the most stringent upper limit on the invisible branching ratio
is\cite{Mitsui,oPS},
\begin{equation}
Br({\rm o-PS}\rightarrow {\rm ``invisible"})\ <\ 2.8\cdot 10^{-6}\ \ \ .
\label{eq:Brinv}
\end{equation}
Clearly, given the remark in (\ref{eq:reldisc}), these upper
bounds exclude the possibility of explaining the o-PS problem in terms
of such decay modes only.

Confronted with this difficulty, theorists have returned 
back\cite{oPS3} to their
calculations trying to establish the next correction\footnote{It
may be shown\cite{Lepage,oPS} that $5\gamma$, $7\gamma$, etc ..., decay modes
of o-PS cannot explain the discrepancy either.} in $(\alpha/\pi)^2$
to the perturbative expansions\cite{oPSth,Khri,oPS} 
used to determine the theoretical
result in (\ref{eq:oPSth}),
\begin{equation}
\lambda^{\rm QED}_{\rm T}({\rm PS}\rightarrow 3\gamma)=
\frac{\alpha^6 m_ec^2}{\hbar}\,
\frac{2(\pi^2-9)}{9\pi}\,\Big[\,1+
(-10.282\pm 0.003)\frac{\alpha}{\pi}+\frac{1}{3}\alpha^2\ln\alpha+
C\left(\frac{\alpha}{\pi}\right)^2\,+\cdots\,\Big]\ \ \ .
\label{eq:lTQED}
\end{equation}
However, in order to explain the observed discrepancy completely
in terms of the next order calculation only, a coefficient $C$ 
of the order of $C\simeq 250\pm 40$ is required\cite{oPS}. Even though 
such a large coefficient
is not to be excluded necessarily,
such a value for the $(\alpha/\pi)^2$ correction is
difficult to contemplate. Incidentally, note that
a value of $C\simeq 40$ only is required by the recent
new experimental result\cite{Asai} given in (\ref{eq:Asai}).

Curiously enough, although the above discrepancy seems to affect only
o-PS decay, none of the suggested mechanisms has
tried to exploit this possible hint towards an explanation.
One such instance is that of invisible decay modes involving
neutrino-antineutrino pairs. 
Indeed, even when slightly massive, (anti)neutrinos emitted in 
positronium decay would be essentially (right)left-handed, since
this is their chirality in the Standard Model (SM) whether there exists
physics beyond it or not. Therefore, momentum and
angular momentum conservation implies that neutrino decay of p-PS
is strongly suppressed---by a factor $(m_\nu/m_e)^2$---or vanishing altogether
in the Standard Model, whereas even in the SM
o-PS does decay via neutrino
pair emission, albeit with the typical small rate of a weak interaction
process.

Obviously, in spite of the fact that this mechanism does indeed distinguish
the decay of the two positronium hyperfine levels, it is expected
to be far too small to explain the discrepancy for the o-PS rate,
the relative weak interaction contribution having to be of the order
$(G_F m_e^2)^2/\alpha^3\simeq 2.4\cdot 10^{-17}$, $G_F$ being Fermi's
coupling constant.
Nevertheless, the neutrino decay channel open essentially only to
o-PS may be enhanced given additional couplings of
the neutrinos. The possibility explicitly considered in this letter
is that of non vanishing magnetic moments for Dirac neutrinos.
Indeed, $C$-odd o-PS decay may then proceed through a single virtual
photon which couples to the electron-positron pair and decays into a
neutrino pair, while this additional
decay channel is forbidden for the
$C$-even p-PS state by charge conjugation invariance.
Therefore, non vanishing neutrino magnetic moments would
induce invisible decay modes of o-PS, but not of p-PS.
In this way, given the measured o-PS
decay rate, it is possible to establish upper bounds
for neutrino magnetic moments. Such an analysis is the purpose
of this letter\footnote{To the authors' knowledge, the only other
study of PS neutrino decay appears
in Ref.\cite{Bern}. However, that work addressed rather
the radiative neutrino decay mode $PS\rightarrow \nu\overline{\nu}\gamma$
only for massless neutrinos in the SM, and did not consider the possibility
of non vanishing magnetic moments.}.

Present experimental upper limits on neutrino magnetic moments
are\footnote{Even though astrophysical or cosmological 
constraints lead to more stringent upper
bounds\cite{PDG,Bernabeu}, such limits are
model dependent and are not included here.} as follows\cite{PDG},
\begin{equation}
\mu_{\nu_e}\,<\,1.08\cdot 10^{-9}\ \mu_B\ \ \ ,\ \ \
\mu_{\nu_\mu}\,<\,7.4\cdot 10^{-10}\ \mu_B\ \ \ ,\ \ \
\mu_{\nu_\tau}\,<\,5.4\cdot 10^{-7}\ \mu_B\ \ \ ,
\label{eq:mulimits}
\end{equation}
$\mu_B$ being the Bohr magneton,
$\mu_B=e\hbar/2m_e$.
Let us also recall the value\cite{Shrock,PDG} of the
magnetic moment of a massive Dirac neutrino in the SM,
\begin{equation}
\mu_\nu\,=\,\frac{3e G_F m_e m_\nu}{8\pi^2 \sqrt{2}}\,=\,
3.2\cdot 10^{-19}\,\left(\frac{m_\nu c^2}{1\ {\rm eV}}\right)\,\mu_B\ \ \ .
\end{equation}

\noindent 2. Let us first consider neutrino decay of positronium
within the SM alone. For the purpose of applications beyond the SM however,
neutrinos are taken to have a non vanishing Dirac mass already,
but the possibility of flavour mixing will be ignored in this paper.
The other assumptions entering the analysis are, on the
one hand, that the couplings
of these massive neutrinos to the $W$ and $Z$ are the usual ones as given
by the SM, and on the other hand, that possible neutral Higgs
contributions are not included since they are expected to be extremely
small for light neutrinos. Relative to $W$ and $Z$ exchange
diagrams, the neutral Higgs exchange amplitude is reduced by a factor
$m_e m_\nu/m^2_h$, a very small ratio indeed. Finally, two further
approximations are effected; on the one hand, the Ps binding energy
is neglected compared to the $e^-e^+$ total rest-mass energy $2m_ec^2$,
namely the electron and positron are taken to annihilate at rest,
and on the other hand, products of ratios of the electron or neutrino masses
to the $W$ and $Z$ masses which appear in $W$ and $Z$ propagators
are taken to be negligible as compared to unity.

These assumptions having been stated, only two amplitudes may contribute
to the neutrino decay process. On the one hand, there is the $Z$ exchange
diagram which contributes for all neutrino flavours\footnote{For the sake
of the analysis, three neutrino flavours whose mass is less
than the electron mass are assumed to be involved in the decay process.
Obviously, the argument would go through for an arbitrary
number of neutrino flavours allowed by the decay kinematics.}
$\nu_\ell$ $(\ell=e,\mu,\tau)$. On the other
hand, there is the charged current amplitude which contributes only
to the electron flavour neutrino channel $\nu_e$ via $W$ exchange.
A straightforward calculation then leads to the following
contributions to the total decay rates. For the singlet state,
one finds,
\begin{equation}
\lambda^{(ZZ)}_S(\nu_\ell)=\frac{\alpha^3}{16}\,
\left(\frac{m^2_eG_F}{\pi\sqrt{2}}\right)^2\,\frac{m_ec^2}{\hbar}\,
\sqrt{1-\frac{m^2_{\nu_\ell}}{m^2_e}}\ \frac{m^2_{\nu_\ell}}{m^2_e}\ \ \ ,
\ \ \ \ell=e,\mu,\tau\ \ \ ,
\end{equation}
\begin{equation}
\lambda^{(WW)}_S(\nu_e)=\frac{\alpha^3}{4}\,
\left(\frac{m^2_eG_F}{\pi\sqrt{2}}\right)^2\,\frac{m_ec^2}{\hbar}\,
\sqrt{1-\frac{m^2_{\nu_e}}{m^2_e}}\ \frac{m^2_{\nu_e}}{m^2_e}\ \ \ ,
\end{equation}
and,
\begin{equation}
\lambda^{(ZW)}_S(\nu_e)=\frac{\alpha^3}{4}\,
\left(\frac{m^2_eG_F}{\pi\sqrt{2}}\right)^2\,\frac{m_ec^2}{\hbar}\,
\sqrt{1-\frac{m^2_{\nu_e}}{m^2_e}}\ \frac{m^2_{\nu_e}}{m^2_e}\ \ \ ,
\end{equation}
where this last contribution follows from the interference of the
$W$ and $Z$ exchange diagrams.

Similarly for the triplet state, one finds,
\begin{equation}
\lambda^{(ZZ)}_T(\nu_\ell)=\frac{\alpha^3}{12}\,
\left(\frac{m^2_eG_F}{\pi\sqrt{2}}\right)^2\,\frac{m_ec^2}{\hbar}\,
\sqrt{1-\frac{m^2_{\nu_\ell}}{m^2_e}}\ 
\left(\,1-\frac{1}{4}\frac{m^2_{\nu_\ell}}{m^2_e}\right)\,
\left(\,1-4\sin^2\theta_W\,\right)^2\ \ ,
\ \ \ell=e,\mu,\tau\ \ \ ,
\end{equation}
\begin{equation}
\lambda^{(WW)}_T(\nu_e)=\frac{\alpha^3}{3}\,
\left(\frac{m^2_eG_F}{\pi\sqrt{2}}\right)^2\,\frac{m_ec^2}{\hbar}\,
\sqrt{1-\frac{m^2_{\nu_e}}{m^2_e}}\ 
\left(\,1-\frac{1}{4}\frac{m^2_{\nu_e}}{m^2_e}\right)\ \ \ ,
\end{equation}
and for the interference contribution,
\begin{equation}
\lambda^{(ZW)}_T(\nu_\ell)=\frac{\alpha^3}{3}\,
\left(\frac{m^2_eG_F}{\pi\sqrt{2}}\right)^2\,\frac{m_ec^2}{\hbar}\,
\sqrt{1-\frac{m^2_{\nu_e}}{m^2_e}}\ 
\left(\,1-\frac{1}{4}\frac{m^2_{\nu_e}}{m^2_e}\right)\,
\left(\,1-4\sin^2\theta_W\,\right)\ \ \ ,
\end{equation}
where $\theta_W$ is the usual weak mixing angle for neutral currents.

A few remarks are in order. First, as was expected for reasons of
the (right)left-handed chirality of (anti)neutrino couplings both
to charged and to neutral currents, contributions to the
singlet decay rate are all suppressed by a factor $m^2_\nu/m^2_e$.
Second, even for the triplet state for which this chiral
suppression is not effective,
$Z$ exchange contributions involve the factor $(1-4\sin^2\theta_W=0.0724)$
leading nevertheless to some suppression as well.
And third, the relevant factor involving the weak coupling constant is
$(m^2_e G_F/\pi\sqrt{2})^2=4.7\cdot 10^{-25}$, 
to be compared to the factors
$\alpha^2=5.33\cdot 10^{-5}$ and 
$(2(\pi^2-9)\alpha^3/9\pi)=2.39\cdot 10^{-8}$ 
relevant to the $2\gamma$ and $3\gamma$
decays of p-PS and o-PS, respectively. Consequently, neutrino
decay of positronium in the SM is very much suppressed, beyond
the reach of any experiment at present. For the sake of the illustration,
in the case of a massless neutrino $\nu_\ell$ ($\ell\ne e$)
one finds for example 
$1/\lambda^{(ZZ)}_T(\nu_\ell, \ell\ne e)
\simeq 5.14\cdot 10^5\ {\rm years}$,
to be compared to $1/\lambda^{\rm QED}_{T}(3\gamma)\simeq
142\ {\rm ns}$!
Incidentally, note that factors of the form $(1-a\,m^2_\nu/m^2_e)$ with
$(a\simeq 1)$, do not
differ significantly from unity for neutrino masses less than say,
a fifth of the electron mass.

3. Let us now extend the above analysis to include the possibility
of neutrino ma\-gne\-tic moments. As is well known\cite{Mohapatra},
this requires massive neutrinos of the Dirac type, hence necessarily
new physics beyond that of the Standard Model. The effective
magnetic moment coupling of Dirac neutrinos is of the form,
\begin{equation}
\overline{\psi_\nu}\,\frac{1}{2}\,
\mu_\nu\,\sigma^{\alpha\beta}\,F_{\alpha\beta}\,\psi_\nu\ \ \ ,
\label{eq:magcoupling}
\end{equation}
where $F_{\alpha\beta}$ is the usual electromagnetic field strength
$F_{\alpha\beta}=\partial_\alpha A_\beta-\partial_\beta A_\alpha$, 
and $\mu_\nu$
the neutrino (anomalous) magnetic moment\cite{Itz}. 
Given this coupling, positronium
decay then proceeds via one more amplitude in addition to those
above, namely through the single photon annihilation channel
$e^-e^+\rightarrow \gamma\rightarrow \nu_\ell\overline{\nu_\ell}$.
In the decay rate, this photon amplitude also interferes with the
charged and neutral current amplitudes of the Standard Model.

For the singlet state, one then finds the identically
vanishing contributions,
\begin{equation}
\lambda^{(\gamma\gamma)}_S(\nu_\ell)=0\ \ \ ,\ \ \ \ell=e,\mu,\tau\ \ \ ,
\end{equation}
\begin{equation}
\lambda^{(\gamma Z)}_S(\nu_\ell)=0\ \ \ ,\ \ \ \ell=e,\mu,\tau\ \ \ ,
\end{equation}
\begin{equation}
\lambda^{(\gamma W)}_S(\nu_e)=0\ \ \ ,
\end{equation}
as is indeed required by charge conjugation invariance of electromagnetic
couplings. On the other hand for the triplet state, one obtains,
\begin{equation}
\lambda^{(\gamma\gamma)}_T(\nu_\ell)=\frac{\alpha^3}{12}\,
\left(\alpha\,\frac{\mu_{\nu_\ell}}{\mu_B}\right)^2\,\frac{m_ec^2}{\hbar}\,
\sqrt{1-\frac{m^2_{\nu_\ell}}{m^2_e}}\ 
\left(\,1+2\frac{m^2_{\nu_\ell}}{m^2_e}\,\right)\ \ \ ,
\ \ \ \ell=e,\mu,\tau\ \ \ ,
\end{equation}
\begin{equation}
\lambda^{(\gamma Z)}_T(\nu_\ell)=\frac{\alpha^3}{4}\,
\left(\frac{m^2_e G_F}{\pi\sqrt{2}}\right)\,
\left(\alpha\,\frac{\mu_{\nu_\ell}}{\mu_B}\right)\,\frac{m_ec^2}{\hbar}\,
\sqrt{1-\frac{m^2_{\nu_\ell}}{m^2_e}}\ 
\frac{m_{\nu_\ell}}{m_e}\
\left(\,1-4\sin^2\theta_W\,\right)\ \ \ ,
\ \ \ \ell=e,\mu,\tau\ \ \ ,
\end{equation}
and
\begin{equation}
\lambda^{(\gamma W)}_T(\nu_e)=\frac{\alpha^3}{2}\,
\left(\frac{m^2_e G_F}{\pi\sqrt{2}}\right)\,
\left(\alpha\,\frac{\mu_{\nu_e}}{\mu_B}\right)\,\frac{m_ec^2}{\hbar}\,
\sqrt{1-\frac{m^2_{\nu_e}}{m^2_e}}\ 
\frac{m_{\nu_e}}{m_e}\ \ \ .
\end{equation}
In these expressions, $\mu_B=e\hbar/2m_e$ is the Bohr magneton.
Note that the $(\gamma Z)$ and $(\gamma W)$ interference contributions
involve directly the ratio $m_\nu/m_e$, in contradistinction
to the pure $(\gamma\gamma)$ contribution. The reason for this result
is that the magnetic neutrino coupling implies a spin flip whereas
the $W$ and $Z$ (anti)neutrino couplings are purely (right)left-handed;
angular momentum conservation thus requires one insertion of the
neutrino mass vertex operator $m_\nu\,\overline{\psi_\nu}\psi_\nu$.
Hence, given small neutrino masses as compared to the electron mass,
it is essentially the $(\gamma\gamma)$ contribution which dominates
over all other magnetic moment contributions, provided the factor
$\alpha\mu_\nu/\mu_B$ is of the order of the effective weak coupling
$(m^2_e G_F/\pi\sqrt{2})$ or larger.

4. Given the above results, the total decay rates
for both singlet and triplet
states into neutrino flavours are
obtained as, respectively,
\begin{equation}
\lambda_S({\rm p-PS}\rightarrow \nu\overline{\nu})=\sum_{\ell=e,\mu,\tau}
\,\lambda^{(Z Z)}_S(\nu_\ell)\ +
\Big[\,\lambda^{(W W)}_S(\nu_e)+\lambda^{(ZW)}_S(\nu_e)\,\Big]\ \ \ ,
\end{equation}
and,
\begin{equation}
\begin{array}{c c l}
\lambda_T({\rm o-PS}\rightarrow \nu\overline{\nu})&=&
\sum_{\ell=e,\mu,\tau}\Big[\,
\lambda^{(\gamma\gamma)}_T(\nu_\ell)+
\lambda^{(\gamma Z)}_T(\nu_\ell)+
\lambda^{(Z Z)}_T(\nu_\ell)\,\Big]\ +\\ \\ 
& &+\ \Big[\,\lambda^{(\gamma W)}_T(\nu_e)+
\lambda^{(W W)}_T(\nu_e)+\lambda^{(ZW)}_T(\nu_e)\,\Big]\ \ \ .
\end{array}
\end{equation}
In particular, in the limit of vanishing neutrino masses, these
expressions reduce to,
\begin{equation}
\lambda_S({\rm p-PS}\rightarrow \nu\overline{\nu})=0\ \ \ ,
\end{equation}
and,
\begin{displaymath}
\lambda_T({\rm o-PS}\rightarrow \nu\overline{\nu})=
\frac{\alpha^3}{12}\,\frac{m_ec^2}{\hbar}\,\sum_{\ell=e,\mu,\tau}
\left(\alpha\frac{\mu_{\nu_\ell}}{\mu_B}\right)^2\ +\\ \\
\end{displaymath}
\begin{equation}
+\ \frac{\alpha^3}{3}
\left(\frac{m^2_e G_F}{\pi\sqrt{2}}\right)^2\,\frac{m_ec^2}{\hbar}\,
\Big[\,1+(1-4\sin^2\theta_W)+\frac{1}{4}N_\nu (1-4\sin^2\theta_W)^2\,\Big]
\ \ \ ,
\label{eq:lTtotal}
\end{equation}
where $N_\nu=3$ is the number of light neutrinos.
Note that these two expressions confirm the announced result,
namely the fact that in practice neutrino disintegration of positronium is
a decay channel open essentially only to the triplet hyperfine state.

The SM contribution from $W$ and $Z$ exchange relative to the
QED $3\gamma$ decay rate being $7.2\cdot 10^{-17}$, let us first
assume that the neutrino magnetic moment contributions
to $\lambda_T({\rm o-PS}\rightarrow \nu\overline{\nu})$ are
dominant, namely let us restrict the discussion now to the
first term only in (\ref{eq:lTtotal}). Compared to the
QED decay rate $\lambda_T^{\rm QED}({\rm PS}\rightarrow 3\gamma)$
in (\ref{eq:lTQED}), the quantity relevant for the confrontation
with experimental results is thus,
\begin{equation}
\frac{\frac{\alpha^3}{12}\frac{m_ec^2}{\hbar}
\sum_{\ell=e,\mu,\tau}\left(\alpha\frac{\mu_{\nu_\ell}}{\mu_B}\right)^2}
{\lambda^{\rm QED}_T({\rm PS}\rightarrow 3\gamma)}\ =\
3.6\cdot 10^6\ \sum_{\ell=e,\mu,\tau}
\left(\alpha\,\frac{\mu_{\nu_\ell}}{\mu_B}\right)^2\ \ \ .
\end{equation}

Taken at face value, if this ratio were to explain completely the
discrepancy expressed in relative terms in (\ref{eq:reldisc}),
one would require the following limit,
\begin{equation}
\sqrt{\sum_{\ell=e,\mu,\tau}\mu^2_{\nu_\ell}}\ \stackrel{<}{_\sim}
3\cdot 10^{-3}\,\mu_B\ \ \ .
\end{equation}
On the other hand, if the corresponding limit stemming from the more
recent measurement in (\ref{eq:Asai}) is used in the same manner,
one derives,
\begin{equation}
\sqrt{\sum_{\ell=e,\mu,\tau}\mu^2_{\nu_\ell}}\ \stackrel{<}{_\sim}
10^{-3}\,\mu_B\ \ \ .
\end{equation}
However, since the upper limit on the invisible branching ratio
in (\ref{eq:Brinv}) already excludes such possibilities,
it is more sensible to use the constraint on this mode to set the
upper bound,
\begin{equation}
\sqrt{\sum_{\ell=e,\mu,\tau}\mu^2_{\nu_\ell}}\ \stackrel{<}{_\sim}
1.2\cdot 10^{-4}\,\mu_B\ \ \ .
\end{equation}
Clearly, these numbers are not competitive with the
experimental limits on neutrino ma\-gne\-tic moments in (\ref{eq:mulimits}).
Even for the least stringent upper bound which applies
in the case of the neutrino $\nu_\tau$, 
$\mu_{\nu_\tau} < 5.4\cdot 10^{-7}\mu_B$, the corresponding branching
ratio is already,
\begin{equation}
Br({\rm o-PS}\rightarrow \nu_\tau\overline{\nu_\tau})\
<\ 5.5\cdot 10^{-11}\ \ \ ,
\end{equation}
namely much less than present experimental upper limits on
branching ratios for invisible decay modes or for the relative
experimental deviations from the QED prediction.

As a matter of fact, it is instructive to consider the situation
when the magnetic moment contribution to the neutrino
branching ratio is comparable to that of the charged and neutral currents,
namely when
$\alpha\mu_\nu/\mu_B=m^2_e G_F/\pi\sqrt{2}$.
Indeed, the corresponding magnetic moment value,
\begin{equation}
\mu_\nu\ =\ 9.39\cdot 10^{-11}\ \mu_B\ \ \ ,
\label{eq:muequal}
\end{equation}
is of the same order of magnitude as the best experimental upper limit 
in (\ref{eq:mulimits}) established for the muon neutrino.
{\sl A posteriori\/}, this is not surprising, since these experimental
limits are determined from neutrino scattering experiments designed
to be sensitive to processes whose strength is typical of weak
interactions. Hence, present limits on neutrino magnetic moments
necessarily correspond to values which render their contributions
comparable to those of the ordinary charged and neutral electroweak currents.
In the present case, the magnetic moment value
in (\ref{eq:muequal}) for a single neutrino contributes the invisible
branching ratio the quantity $1.7\cdot 10^{-18}$, which is indeed
close to the neutrino branching ratio in the SM of
$7.2\cdot 10^{-18}$, 
namely the contribution of the second term in (\ref{eq:lTtotal}).

In fact, using the upper limits in (\ref{eq:mulimits}) as values
for the neutrino magnetic moments together with the result in
(\ref{eq:lTtotal}), one obtains the following total neutrino
branching ratio,
\begin{equation}
Br({\rm o-PS}\rightarrow \nu\overline{\nu})\ =\ 5.6\cdot 10^{-11}\ \ \ ,
\label{eq:Brlimits}
\end{equation}
which is thus dominated by the magnetic moment
contribution of the $\nu_\tau$ neutrino.

However, the value in (\ref{eq:Brlimits}) is larger than upper bounds
recently es\-ta\-bli\-shed\cite{Masso} for the
branching ratio of ``exotic" positronium
decays on the basis of primordial
nucleosynthesis. In fact, since
the electroweak charged and neutral current contributions relevant
to the thermal equilibrium of three light neutrinos are already
included\cite{Weinberg} in the standard cosmological model, 
the ``exotic" contributions to be considered
here are solely those stemming from the neutrino magnetic moments. 
Moreover, the associated
couplings (\ref{eq:magcoupling}) being dimension five operators,
it is actually the upper bound of $2\cdot 10^{-17}$ on the
branching ratio, associated
to case B) of Ref.\cite{Masso}, which is relevant in our case
and which thus applies only to the total contribution
of neutrino magnetic moments. 
Correspondingly, this upper limit implies,
\begin{equation}
\sqrt{\sum_{\ell=e,\mu,\tau}\mu^2_{\nu_\ell}}\ \stackrel{<}{_\sim}
3.2\cdot 10^{-10}\,\mu_B\ \ \ .
\end{equation}
Incidentally, note that
this upper bound value is also typical of astrophysical
constraints on neutrino magnetic moments\cite{PDG}.

5. In conclusion, even though neutrino decay of positronium in the
Standard Model and beyond it does indeed contribute to invisible
decay modes of the triplet orthopositronium state, but practically
not to the decay of the singlet parapositronium state, present experimental
limits on neutrino magnetic moments imply that such processes
cannot provide even for a partial contribution towards a resolution
of the possible orthopositronium lifetime puzzle\cite{oPS},
which however, may have been resolved by a recent new measurement\cite{Asai}.
Moreover, these limits also establish
that the observation of such decay modes is far beyond present
experimental capabilities. Nevertheless, when considered
together with recent arguments\cite{Masso} 
based on primordial nucleosynthesis, the orthopositronium neutrino
decay mode implies upper bounds on neutrino magnetic moments which
are more stringent than present experimental limits.

Although the analysis assumes massive Dirac neutrinos without
flavour mixing, it should be clear that similar
conclusions would apply more generally for transition magnetic
moments as well, whether the massive neutrinos are Dirac or
Majorana spinors with flavour mixing.

Note also that an analogous discussion could be developed were neutrinos
to carry electric charges\cite{Joshi} $eQ_\nu$.
Under such circumstances, the associated couplings would imply
contributions to positronium decay such as those above
in which the ratio $\alpha\mu_\nu/\mu_B$ is replaced by the
factor $\alpha Q_\nu$, thereby leading to
similar conclusions. For example in the case of the triplet
decay rate\footnote{Here again,
contributions to the singlet rate vanish by charge conjugation
invariance.}, the modulus squared photon amplitude leads to,
\begin{equation}
\lambda^{(\gamma\gamma)}_T(\nu_\ell;Q_{\nu_\ell})=
\frac{1}{6}\,\alpha^3\,(\alpha Q_{\nu_\ell})^2\,
\frac{m_ec^2}{\hbar}\,
\sqrt{1-\frac{m^2_{\nu_\ell}}{m^2_e}}\
\left(1+\frac{1}{2}\frac{m^2_{\nu_\ell}}{m^2_e}\right)\ \ \ ,\ \ \
\ell=e,\mu,\tau\ \ \ .
\end{equation}
Therefore, the same types of upper bounds as those derived
above for the ratios $\mu_\nu/\mu_B$ would apply to
the neutrino electric charges $Q_\nu$. 
However, these constraints
are not competitive with existing limits\cite{Joshi}.
In this respect, it is of interest
to remark that based on the anisotropy
of the microwave background, an upper bound of
$|Q_\nu| < 4.8\cdot 10^{-34}$ was recently
obtained\cite{Sengupta}, which is thus
many orders of magnitude more stringent
than any existing limit on neutrino magnetic moments.

\vspace{20pt}

It is a pleasure to thank Prof. Jean Pestieau for his interest
in this work and for discussions.

\end{document}